\documentclass[
 amsmath,amssymb,
 aps, prd, physrev,
]{revtex4-2}
\usepackage{graphicx}
\usepackage{dcolumn}
\usepackage{bm}
\usepackage{xcolor}
\usepackage{ulem}
\usepackage{cancel}

\begin{document}

\title{\textbf{Multiplicity-dependent Hadron Enhancement in High-Energy $pp$ and $p$--Pb Collisions within an Effective Mass-Scale Framework}}
\author{R.~C.~Baral}
\email{rcbaral@ravenshawuniversity.ac.in}
\affiliation{Department of Physics, Ravenshaw University, Cuttack, India}

\author{B.~Mohanty}
\affiliation{National Institute of Science Education and Research, Jatni 752050, Odisha, India}
\affiliation{Homi Bhabha National Institute, Training School Complex, Anushakti Nagar, Mumbai 400094, Maharashtra, India}

\begin{abstract}

  The multiplicity dependence of identified hadron yield ratios in high-energy $pp$ and $p$--Pb collisions has commonly been interpreted in terms of strangeness-driven scaling and canonical suppression effects. In this work, we investigate whether the observed enhancement hierarchy may also admit a complementary phenomenological organization involving effective hadronic and valence-quark mass scales relative to a pion baseline. A simultaneous description of non-strange, strange, and multi-strange hadron-to-pion ratios is performed for $pp$ collisions at $\sqrt{s}=7$ TeV and $p$--Pb collisions at $\sqrt{s_{\mathrm{NN}}}=5.02$ TeV using an effective mass-scale parametrization. The stability of the parametrization is tested through reduced $\chi^{2}$ values, pull distributions, parameter correlations, information-criterion comparisons, cross-system predictions, multiplicity-range variations, and studies of observables not included in the fit. Additional investigations involving relative enhancement patterns and hidden-strangeness $\phi$ mesons are used to examine the extent to which the observed hierarchy is uniquely characterized by simple open-strangeness ordering. The analysis indicates that the multiplicity dependence of identified hadron production can be organized phenomenologically through an interplay of open strangeness, hadron species dependence, hidden-strangeness structure, and effective mass-related scales. The present framework should be interpreted as a complementary phenomenological description rather than as a microscopic theory of hadron production.
\end{abstract}

\maketitle


\section{Introduction}

The study of identified hadron production in high-energy hadronic collisions provides essential insight into the mechanisms of particle production and the properties of the underlying medium. In nucleus--nucleus collisions, the enhancement of strange and multi-strange hadrons relative to non-strange species has long been considered a signature of the formation of a deconfined quark--gluon plasma (QGP)~\cite{Rafelski1982,Koch1986}. Remarkably, similar features have been observed in high-multiplicity proton--proton ($pp$) and proton--nucleus ($p$--Pb) collisions at the Large Hadron Collider (LHC), where the relative production of strange hadrons increases with event multiplicity~\cite{ALICE2017Nature,ALICE2013pPb,ALICE2015pp,ALICE2019multi}. This observation has motivated extensive theoretical and phenomenological investigations. In the statistical hadronization framework, particle yields are described assuming thermal equilibrium at chemical freeze-out~\cite{BraunMunzinger2003,Andronic2018}. In small systems, canonical suppression of strangeness has been proposed as a mechanism to explain the multiplicity dependence, where conservation laws restrict strange quark production in low-multiplicity events~\cite{Hamieh2000,Becattini2008}. Alternative approaches include microscopic models based on string dynamics, color reconnection, and rope formation~\cite{Bierlich2015,Bierlich2018}, as well as quark recombination or coalescence mechanisms that become relevant in dense partonic environments~\cite{Fries2003,Greco2003}. Experimental measurements by the ALICE Collaboration have established a clear hierarchy in the enhancement pattern, where the increase in yield ratios scales with the strangeness content of the hadron~\cite{ALICE2017Nature}. This behavior has been successfully described using empirical parametrizations that explicitly depend on the number of strange valence quarks~\cite{ALICE2017Nature}.

Although these approaches differ substantially in their underlying microscopic assumptions, they share the common feature that the observed multiplicity dependence emerges through correlations between hadron production and intrinsic hadronic properties such as quark content, mass scale, and production dynamics. In statistical and hadron resonance gas (HRG) based descriptions, the enhancement hierarchy is associated with the gradual reduction of canonical suppression effects with increasing system size and particle multiplicity~\cite{Hamieh2000,Becattini2008,BraunMunzinger2003,Andronic2018}. In microscopic event generators such as \textsc{PYTHIA}, mechanisms including color reconnection and rope hadronization generate collective string effects that can enhance the production of heavier and strange hadrons in high-multiplicity environments~\cite{Bierlich2015,Bierlich2018}. Similarly, coalescence and recombination models relate the observed hadron yields to effective partonic degrees of freedom and quark-level recombination probabilities in dense systems~\cite{Fries2003,Greco2003}.

However, strangeness is intrinsically correlated with other hadronic properties, particularly the hadron mass and the effective mass scale associated with its valence quark content. Consequently, it remains worthwhile to investigate whether the experimentally observed enhancement hierarchy may also admit a broader phenomenological organization involving effective hadronic and quark mass scales, in addition to conventional strangeness-based interpretations. In this context, the observed multiplicity dependence may reflect an interplay between partonic degrees of freedom, hadron formation dynamics, and effective mass-related scales that are not uniquely characterized by strangeness quantum numbers alone.

In this work, we investigate the multiplicity dependence of identified hadron yield ratios in $pp$ collisions at $\sqrt{s}=7$~TeV and $p$--Pb collisions at $\sqrt{s_{\mathrm{NN}}}=5.02$~TeV within a phenomenological scaling framework. Motivated by the observed correlation between enhancement patterns and effective mass-related quantities, we construct an effective scaling parametrization based on effective hadronic and quark-level mass-scale differences relative to the pion baseline. A simultaneous description of a broad set of hadron-to-pion ratios, including non-strange, strange, and multi-strange hadrons, is performed and compared with commonly used strangeness-based scaling parametrizations. The present analysis is intended as a phenomenological study of scaling behavior and does not attempt to replace microscopic or thermal descriptions of hadron production.

Beyond reproducing the measured ratios, the proposed parametrization is tested through its predictive consistency for independent observables not included in the fit, including yield ratios with different denominators and resonance production. In addition, complementary studies of relative enhancement and effective mass-scale organization are performed in order to examine the interplay between hadronic structure, valence-quark mass scales, and multiplicity-dependent production patterns. These studies aim to examine whether the multiplicity-dependent hierarchy admits a complementary effective scaling organization involving hadronic and quark mass scales, and to explore the extent to which the observed trends may reflect complementary effective mass-scale organization beyond simple open-strangeness ordering. Special attention is also given to the hidden-strangeness $\phi$ meson as a nontrivial probe of the proposed scaling framework. The physical origin of the observed scaling behavior, and its possible connection to underlying QCD dynamics, remain open questions beyond the scope of the present phenomenological analysis. 

The paper is organized as follows. In Sec.~II, we describe the data sets, parametrization framework, and statistical methodology used in the analysis. Section~III presents the simultaneous fit results, predictive and robustness studies, and investigations of the disentangling between strangeness and effective mass-scale effects. In Sec.~IV, we discuss the physical interpretation and limitations of the proposed phenomenological framework. Finally, conclusions are summarized in Sec.~V.

%
\section{Data Sets and Methodology}

The present analysis is based on measurements of identified hadron production in $pp$ collisions at $\sqrt{s}=7$~TeV and $p$--Pb collisions at $\sqrt{s_{\mathrm{NN}}}=5.02$~TeV performed by the ALICE Collaboration~\cite{ALICE2013pPb,ALICE2015pp,ALICE2019multi}. The study focuses on the multiplicity dependence of integrated hadron yield ratios involving light-flavor and strange hadrons as functions of the charged-particle multiplicity density $\langle dN_{\mathrm{ch}}/d\eta \rangle$ at midrapidity.
The observables included in the simultaneous fit are
\begin{equation}
 p/\pi,\quad K^{0}_{S}/\pi,\quad \Lambda/\pi,\quad \Xi/\pi,\quad \Omega/\pi, \nonumber
\end{equation}
measured over different multiplicity classes in both collision systems. In addition, several independent observables not included in the fit,
\begin{equation}
\Lambda/K^{0}_{S},\quad \Xi/K^{0}_{S},\quad p/K^{0}_{S},\quad \phi/\pi,\quad \Xi^{*}/\pi, \nonumber
\end{equation}
are employed to examine the predictive consistency of the proposed parametrization. For all fits and statistical analyses, the total uncertainty is obtained by combining the statistical and uncorrelated systematic uncertainties in quadrature. Correlated systematic uncertainties are not explicitly included in the $\chi^{2}$ minimization procedure because the full covariance matrices are not available for all measured observables and multiplicity intervals considered in the present analysis. Since the study primarily involves normalized yield ratios measured within the same experimental setup, a significant fraction of the correlated normalization-related systematic uncertainties is expected to cancel. The resulting $\chi^{2}$ values should therefore be interpreted as effective measures of the relative goodness of fit rather than as strict statistical likelihood estimators. The reduced $\chi^{2}$ values obtained in the present analysis are significantly below unity, which likely reflects the combined effect of conservative experimental uncertainties together with the incomplete treatment of correlated systematic effects. Consequently, the extracted $\chi^{2}/\mathrm{ndf}$ values should be interpreted primarily as relative indicators of fit quality and parameter stability rather than as precise statistical likelihood measures.

The multiplicity dependence of the normalized hadron yield ratios is described using the phenomenological parametrization

\begin{eqnarray} \label{eq:parametrization}
   \frac{N/D}{(N/D)^{{pp}}_{\rm{INEL} > 0}}=1+ C_{0} \ {\rm log}\left[ \frac{\langle {\rm d} N_{\rm ch} / {\rm d} {\eta} \rangle}  {\langle {\rm d} N_{\rm ch} / {\rm d} {\eta} \rangle ^{pp}_{\rm INEL > 0}} \right]
    + \frac{C_{1}}{\langle {\rm d} N_{\rm ch} / {\rm d} {\eta} \rangle} \ {\rm log}  \left[ \frac{\langle {\rm d} N_{\rm ch} / {\rm d} {\eta} \rangle}  {\langle {\rm d} N_{\rm ch} / {\rm d} {\eta} \rangle ^{pp}_{\rm INEL > 0}} \right]&{}&, 
\end{eqnarray}
where
\begin{eqnarray}
   \ C_{0} &=& a \left(({M}_{{\it q},\rm N}){^\textit{b}}-({M}_{{\it q},\rm D})^\textit{b}\right), \nonumber \\
  \ C_{1} &=& c \left({M}_{{\it h},\rm N} - {M}_{{\it h},\rm D}\right). \nonumber 
\end{eqnarray}
where $N$ and $D$ denote the hadrons appearing in the numerator and denominator of the corresponding yield ratio, respectively. Here, $M_{h,N}$ and $M_{h,D}$ represent the corresponding hadron masses, while $M_{q,N}$ and $M_{q,D}$ denote the associated effective valence quark mass scales. The hadron masses and current quark masses are taken from the Particle Data Group compilation~\cite{PDG2020}.

The effective valence quark and hadronic mass scale are defined phenomenologically as
\begin{equation}
M_{q} = 2\sum_{i} m_{q,i}, \qquad
M_h = 2M_{\mathrm{hadron}},  \nonumber
\end{equation}
where the summation runs over the valence quarks constituting the hadron. $M_{\mathrm{hadron}}$ denotes the physical hadron mass. The overall factor of two introduced in both the effective quark and hadronic mass scales is phenomenological and is intended to represent the effective particle--antiparticle production threshold relevant for the multiplicity-dependent enhancement hierarchy. In the present parametrization, the current quark masses $m_{u}=2.2~\mathrm{MeV}, \qquad m_{d}=4.7~\mathrm{MeV}, \qquad m_{s}=96~\mathrm{MeV}$ are used throughout the baseline parametrization. The chosen functional form is motivated phenomenologically by the experimentally observed multiplicity evolution of identified hadron yield ratios, which typically exhibits a relatively rapid enhancement at low multiplicity followed by a slower, approximately saturating behavior toward higher multiplicities. The logarithmic multiplicity dependence therefore provides a simple effective description of the smooth enhancement trend, while the inverse multiplicity term allows additional curvature required to describe the stronger low-multiplicity evolution observed particularly for heavier strange hadrons. The additive structure involving effective hadronic and valence-quark mass scales was introduced as a minimal phenomenological ansatz to examine whether both quark-level and hadronic-scale contributions can simultaneously organize the observed multiplicity hierarchy. The parametrization is not derived from a microscopic production model and should therefore be interpreted solely as an effective phenomenological scaling form.

The $\phi$ meson requires separate consideration within the present framework because it is a hidden-strangeness ($s\bar{s}$) state carrying zero net open strangeness. Unlike open-strange hadrons such as $K_S^0$, $\Lambda$, $\Xi$, and $\Omega$, the $\phi$ meson does not involve net strange-quark transport in the same manner. Consequently, the phenomenological open-flavor scaling prescription used for the remaining hadrons is not applied to the $\phi$ meson contribution. The effective quark and hadron mass scales entering Eq.~(\ref{eq:parametrization}) for the $\phi/\pi$ ratio are therefore taken as
\begin{equation}
  M_{q,N}(\phi)=2m_s, \qquad
  M_{h,N}(\phi)=M_{\phi}, \nonumber
\end{equation}
while the pion denominator retains the standard definition used throughout the analysis. The $\phi$ meson is thus examined separately within the same phenomenological scaling framework in order to explore the behavior of the hidden-strangeness sector.

The fit parameters are determined through a global $\chi^{2}$ minimization procedure applied simultaneously to all fitted hadron-to-pion ratios in both collision systems. The $\chi^{2}$ function is defined as
\begin{equation}
\chi^{2} = \sum_{i=1}^{N}
\frac{\left(R_{i}^{\mathrm{data}}-R_{i}^{\mathrm{fit}}\right)^{2}}
{\sigma_{i}^{2}},
\nonumber
\end{equation}
where $R_{i}^{\mathrm{data}}$ and $R_{i}^{\mathrm{fit}}$ denote the measured and parametrized values of the corresponding observable for the $i$th data point, respectively, and $\sigma_i$ represents the associated total uncertainty. Here, $N$ denotes the total number of fitted data points included in the simultaneous analysis. The simultaneous analysis includes a total of $N=80$ fitted data points with $k=3$ free fit parameters, resulting in $\mathrm{ndf}=77$. The quality of the fit is quantified using the reduced chi-square, $\chi^{2}/\mathrm{ndf}$, together with pull distributions defined as
\begin{equation}
\mathrm{Pull}_{i} = \frac{\mathrm{Data}_{i} - \mathrm{Fit}_{i}}{\sigma_{i}}. \nonumber
\end{equation}
\begin{table}[htbp]
\centering
\caption{Best-fit parameters obtained from the simultaneous fit to all hadron-to-pion ratios in $pp$ and $p$--Pb collisions.}
\begin{tabular}{cc}
\hline
Parameter & Value \\
\hline
$a$ & $1.350 \pm 0.134$ \\
$b$ & $1.836 \pm 0.107$ \\
$c$ & $0.181 \pm 0.035$ \\
\hline
$\chi^{2}$ & 29.10 \\
$\chi^{2}/\mathrm{ndf}$ & 0.38 \\
\hline
\end{tabular}
\label{tab:fit_parameters}
\end{table}

The best-fit parameters and fit quality indicators are summarized in Table~\ref{tab:fit_parameters}. The quoted parameter uncertainties are obtained from the covariance matrix of the fit. A strong positive correlation is observed between parameters $a$ and $b$, with correlation coefficient $\rho(a,b)=0.92$, indicating partial coupling between the overall scaling strength and the effective scaling exponent within the present phenomenological parametrization. The parameter correlations obtained from the covariance matrix of the simultaneous fit are summarized in Table~\ref{tab:corr_matrix}.

\begin{table}[htbp]
\centering
\caption{Correlation matrix of the fit parameters obtained from the simultaneous global fit.}
\begin{tabular}{c|ccc}
\hline
 & $a$ & $b$ & $c$ \\
\hline
$a$ & 1.000 & 0.921 & $-0.028$ \\
$b$ & 0.921 & 1.000 & 0.153 \\
$c$ & $-0.028$ & 0.153 & 1.000 \\
\hline
\end{tabular}
\label{tab:corr_matrix}
\end{table}
To further compare the descriptive capability of the present framework with alternative phenomenological parametrizations while accounting for model complexity, information-criterion based measures are also evaluated. To compare the present framework with the commonly used strangeness-based scaling while accounting for the difference in the number of fit parameters, the Akaike Information Criterion (AIC)~\cite{Akaike1974} and Bayesian Information Criterion (BIC)~\cite{Schwarz1978} are also evaluated:
\begin{equation}
\mathrm{AIC} = \chi^{2}+2k, \nonumber
\end{equation}
\begin{equation}
\mathrm{BIC} = \chi^{2}+k\ln(N), \nonumber
\end{equation}
where $k$ denotes the number of fit parameters and $N$ represents the total number of fitted data points.

In addition to the fit analysis, relative enhancement patterns are examined using identified hadron yield ratios measured at different charged-particle multiplicities. The enhancement is defined through
\begin{equation}
E = \frac{R(\mathrm{mult})}{R(\mathrm{low~mult})}, \nonumber
\end{equation}
where $R$ denotes the corresponding hadron yield ratio. The enhancement behavior is further investigated through complementary studies involving hadron mass, open-strangeness structure, and hidden-strangeness $\phi$ mesons using pp collision data at $\sqrt{s}=13$~TeV. These studies are intended to explore possible phenomenological organization underlying the observed multiplicity hierarchy and to examine the extent to which the enhancement trends are uniquely characterized by simple open-strangeness ordering. In addition to the baseline simultaneous fit, several robustness and stability studies are performed, including cross-system predictive tests, leave-one-out analyses, multiplicity-range variations, and alternative hidden-strangeness prescriptions for the $\phi$ meson. These studies are used to examine the stability and predictive capability of the proposed phenomenological framework.

%
%
%
\section{Results and Discussion}

\subsection{Simultaneous fit to hadron-to-pion ratios}

\begin{figure*}[htbp]
\centering
\includegraphics[width=0.5\linewidth]{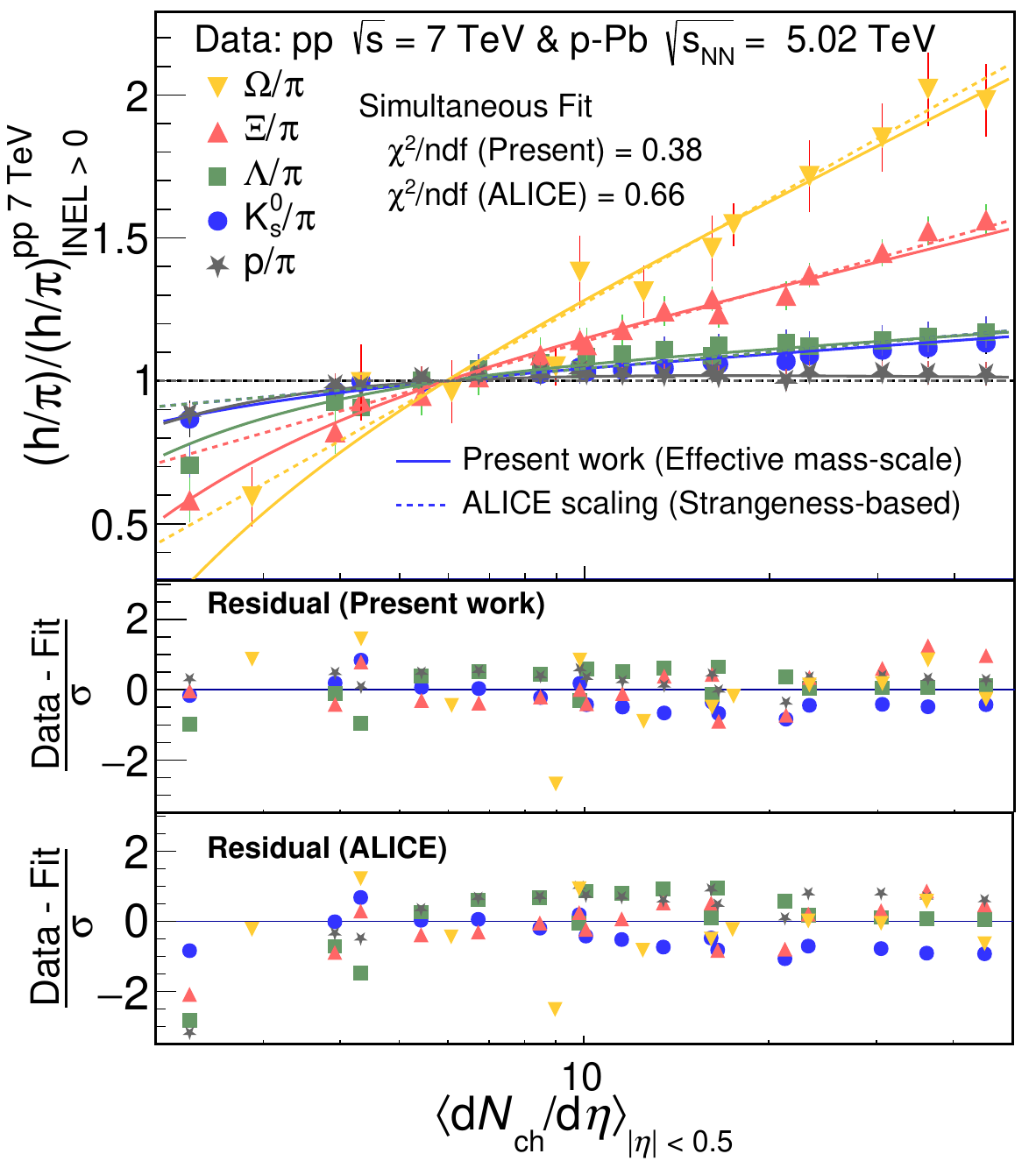}
\caption{Simultaneous description of the normalized hadron-to-pion yield ratios in $pp$ collisions at $\sqrt{s}=7$~TeV and $p$--Pb collisions at $\sqrt{s_{\mathrm{NN}}}=5.02$~TeV using the present effective mass-scale parametrization (solid curves) and the strangeness-based scaling used by the ALICE Collaboration~\cite{ALICE2017Nature} (dashed curves). The upper panel shows the measured yield ratios as functions of charged-particle multiplicity density, while the middle and lower panels display the corresponding pull distributions for the present parametrization and the ALICE scaling, respectively.}
\label{fig:GlobalFit}
\end{figure*}

Figure~\ref{fig:GlobalFit} presents the simultaneous description of the normalized hadron-to-pion yield ratios in $pp$ collisions at $\sqrt{s}=7$~TeV and $p$--Pb collisions at $\sqrt{s_{\mathrm{NN}}}=5.02$~TeV using the parametrization of Eq.~(\ref{eq:parametrization}). The fit includes the multiplicity dependence of $p/\pi,\quad K^{0}_{S}/\pi,\quad \Lambda/\pi,\quad \Xi/\pi,\quad \Omega/\pi$, covering both non-strange and multi-strange hadrons over a broad multiplicity range.

For comparison, the figure also includes the strangeness-based empirical scaling previously used to describe the multiplicity hierarchy observed by the ALICE Collaboration~\cite{ALICE2017Nature}. The upper panel shows the measured ratios together with the corresponding parametrizations, while the lower panels display the pull distributions for the present framework and the strangeness-based scaling, respectively.
The present parametrization provides a consistent simultaneous description of all fitted observables across both collision systems. In particular, the residual distributions exhibit a reduced overall spread and a smaller systematic bias compared to the strangeness-based scaling. The fit yields a mean pull of approximately $0.04$ with an RMS of $0.60$, while the corresponding values for the strangeness-based parametrization are approximately $-0.07$ and $0.85$, respectively. This indicates that the present framework provides a comparatively uniform description of the data across different hadron species and multiplicity intervals within the present phenomenological analysis.

The overall fit quality is further quantified through the reduced chi-square and information criteria. The present parametrization yields a smaller $\chi^{2}/\mathrm{ndf}$ compared to the strangeness-based scaling despite simultaneously describing a larger number of observables, including the $p/\pi$ ratio. The corresponding AIC and BIC values are also found to be smaller for the present framework, indicating that the improved agreement is not solely a consequence of the additional fit parameter. The quantitative comparison summarized in Table~\ref{tab:FitQuality} indicates that the present framework provides a systematically improved description of the data across multiple statistical measures. The stability of the parametrization is further examined through predictive tests, multiplicity-range variations, leave-one-out analyses, and cross-system comparisons discussed in the following subsections.
\begin{table}[t]
\centering
\caption{
Comparison of the statistical quality indicators for the present effective mass-scale parametrization and the strangeness-based scaling used by the ALICE Collaboration~\cite{ALICE2017Nature}. The table includes the reduced chi-square, mean pull, RMS of the pull distribution, and the corresponding AIC and BIC values.
}
\label{tab:FitQuality}
\begin{ruledtabular}
\begin{tabular}{lccccc}
Model & $\chi^{2}/\mathrm{ndf}$ & Mean Pull & RMS & AIC & BIC \\
\hline
Present work & 0.38 & ~~0.04 & 0.61 & 35.3 & 42.4 \\
ALICE scaling & 0.66 & $-0.07$ & 0.85 & 44.3 & 48.5 \\
\end{tabular}
\end{ruledtabular}
\end{table}
An important feature of the present parametrization is that it incorporates both an effective valence quark mass scale and a hadronic mass-scale difference relative to the pion baseline. The first term in Eq.~(\ref{eq:parametrization}) primarily controls the multiplicity-dependent hierarchy among different hadron species, while the second term introduces an additional hadronic mass-scale contribution that remains subleading but non-negligible, particularly for heavier hadrons and low-multiplicity regions. The simultaneous inclusion of these two scales enables the framework to describe both strange and non-strange hadrons within a common phenomenological structure.

Although the observed multiplicity hierarchy has commonly been interpreted in terms of explicit strangeness scaling, the present results indicate that a complementary effective mass-scale organization can also account for a substantial part of the observed behavior within a common phenomenological framework. The analysis therefore suggests that the multiplicity hierarchy may not be uniquely determined by simple open-strangeness ordering alone, but may additionally reflect an interplay between quark-level and hadronic mass scales in hadron production.

\subsection{Predictions for independent hadron yield ratios}

\begin{figure}[t]
\centering
\includegraphics[width=0.5\columnwidth]{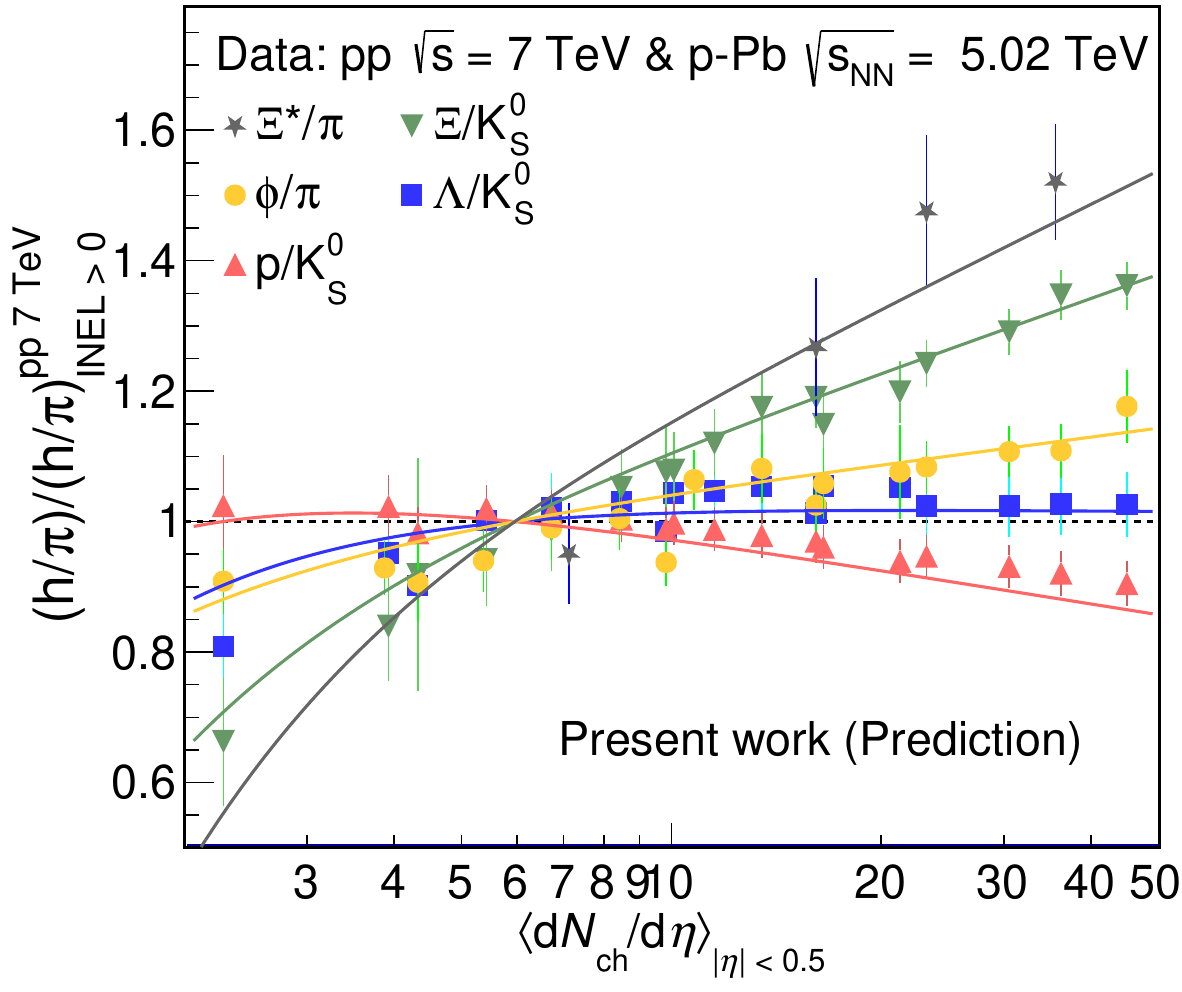}
\caption{
Multiplicity dependence of independent hadron yield ratios not included in the fit, compared with the predictions of the present effective mass-scale parametrization. The figure shows $\Xi/K^{0}_{S}$, $\Lambda/K^{0}_{S}$, $p/K^{0}_{S}$, $\phi/\pi$, and $\Xi^{*}/\pi$ as functions of $\langle dN_{\mathrm{ch}}/d\eta \rangle$. The data are compared with the model curves obtained using the parameters fixed from the simultaneous fit to the normalized hadron-to-pion ratios. For the $\phi/\pi$ observable, the hidden-strangeness prescription discussed in Sec.~II is employed.}
\label{fig:Prediction}
\end{figure}

A central test of any phenomenological parametrization is whether it can describe observables that were not used in the fit. Figure~\ref{fig:Prediction} presents such a test for the present effective mass-scale framework. The parameters of Eq.~(\ref{eq:parametrization}) were fixed solely from the simultaneous fit to the normalized hadron-to-pion ratios discussed in the previous subsection, while the ratios shown here were not included in the fitting procedure. In particular, the figure compares the model predictions with the measured multiplicity dependence of
\begin{equation}
\Xi/K^{0}_{S},\quad \Lambda/K^{0}_{S},\quad p/K^{0}_{S},\quad \phi/\pi,\quad \Xi^{*}/\pi. \nonumber
\end{equation}
These observables probe different aspects of hadron production, including ratios with different denominators, baryon-to-meson combinations, hidden-strangeness production, and resonance formation.

The present parametrization reproduces the measured multiplicity trends of these independent ratios with good overall accuracy. In particular, the ratios involving $K^{0}_{S}$ in the denominator are captured consistently, indicating that the effective mass-scale structure obtained from the $h/\pi$ fit is not restricted to a single normalization choice.
In the present framework, the hidden-strangeness nature of the $\phi$ meson is incorporated through the modified effective mass-scale prescription discussed in Sec.~II. The resulting prediction provides a moderately improved overall agreement of the observed multiplicity dependence compared to a uniform open-flavor treatment, indicating that hidden-strangeness states exhibit nontrivial behavior within the proposed phenomenological organization. Likewise, the agreement for $\Xi^{*}/\pi$ suggests that the parametrization is not limited to ground-state hadrons alone, but retains predictive consistency for resonance production as well. Taken together, these results show that the proposed parametrization captures more general systematics of multiplicity-dependent hadron production than the observables used to determine the fit parameters. This predictive behavior provides additional support for the effective mass-scale interpretation of the observed hadron-production hierarchy.

\subsection{Robustness and stability tests}

In order to examine the stability and predictive consistency of the proposed phenomenological parametrization, several complementary robustness tests are performed. These studies include cross-system predictive tests, variations of the fitted multiplicity interval, and leave-one-out analyses in which selected observables are excluded from the fitting procedure.
\begin{figure}[t]
\centering
\includegraphics[width=0.4\columnwidth]{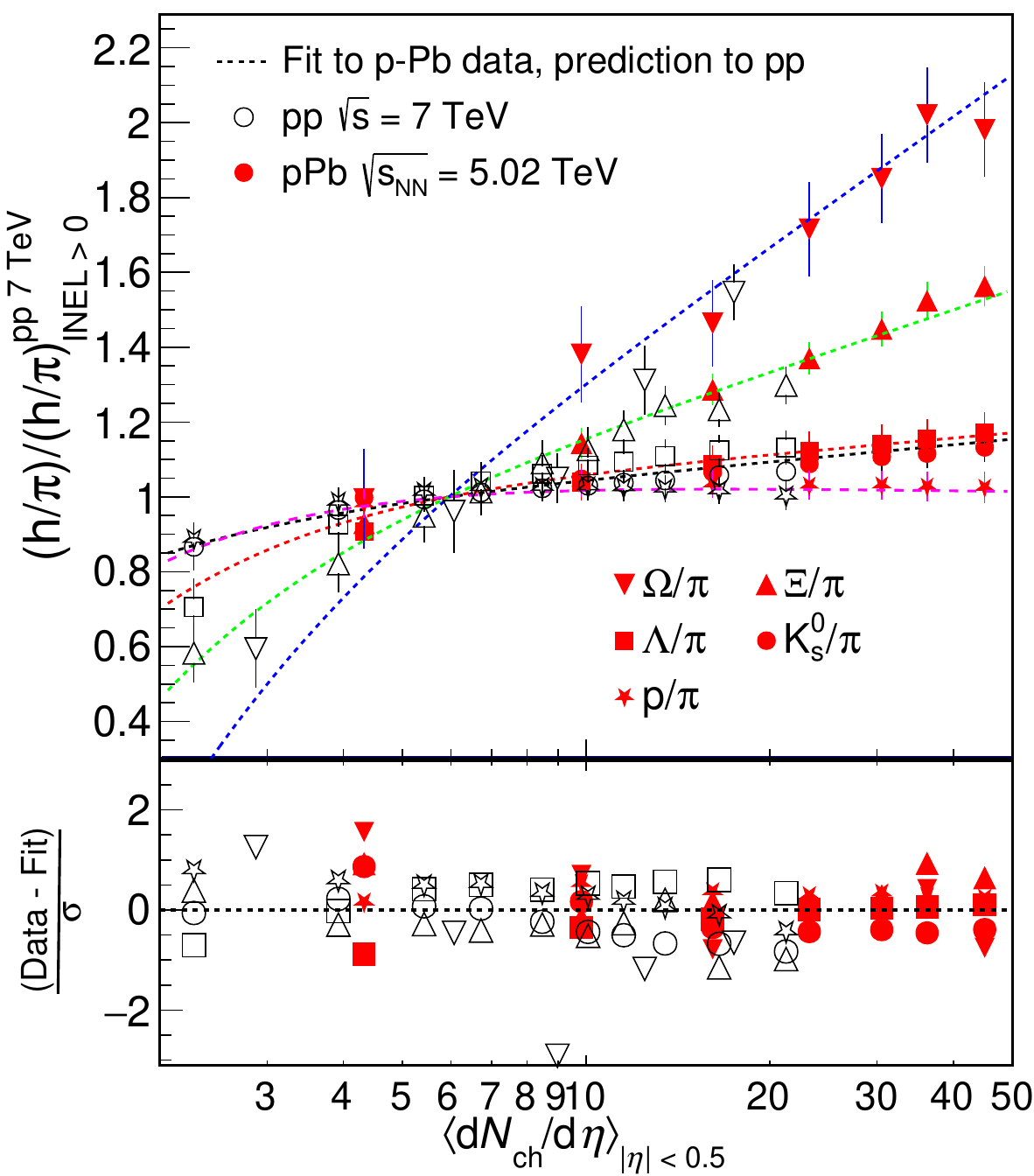}
\caption{
Cross-system predictive test of the present effective mass-scale parametrization. The fit parameters are determined exclusively from the p--Pb data and subsequently used to predict the multiplicity dependence of the corresponding pp hadron-to-pion ratios. The upper panel shows the normalized ratios $(h/\pi)/(h/\pi)_{\mathrm{INEL}>0}^{pp~7~\mathrm{TeV}}$ for $K^{0}_{S}/\pi$, $p/\pi$, $\Lambda/\pi$, $\Xi/\pi$, and $\Omega/\pi$ as functions of $\langle dN_{\mathrm{ch}}/d\eta \rangle$. The lower panel presents the standardized residuals $(\mathrm{Data}-\mathrm{Fit})/\sigma$.}
\label{fig:pPbToPP}
\end{figure}
\begin{figure}[thb]
\centering
\includegraphics[width=0.4\columnwidth]{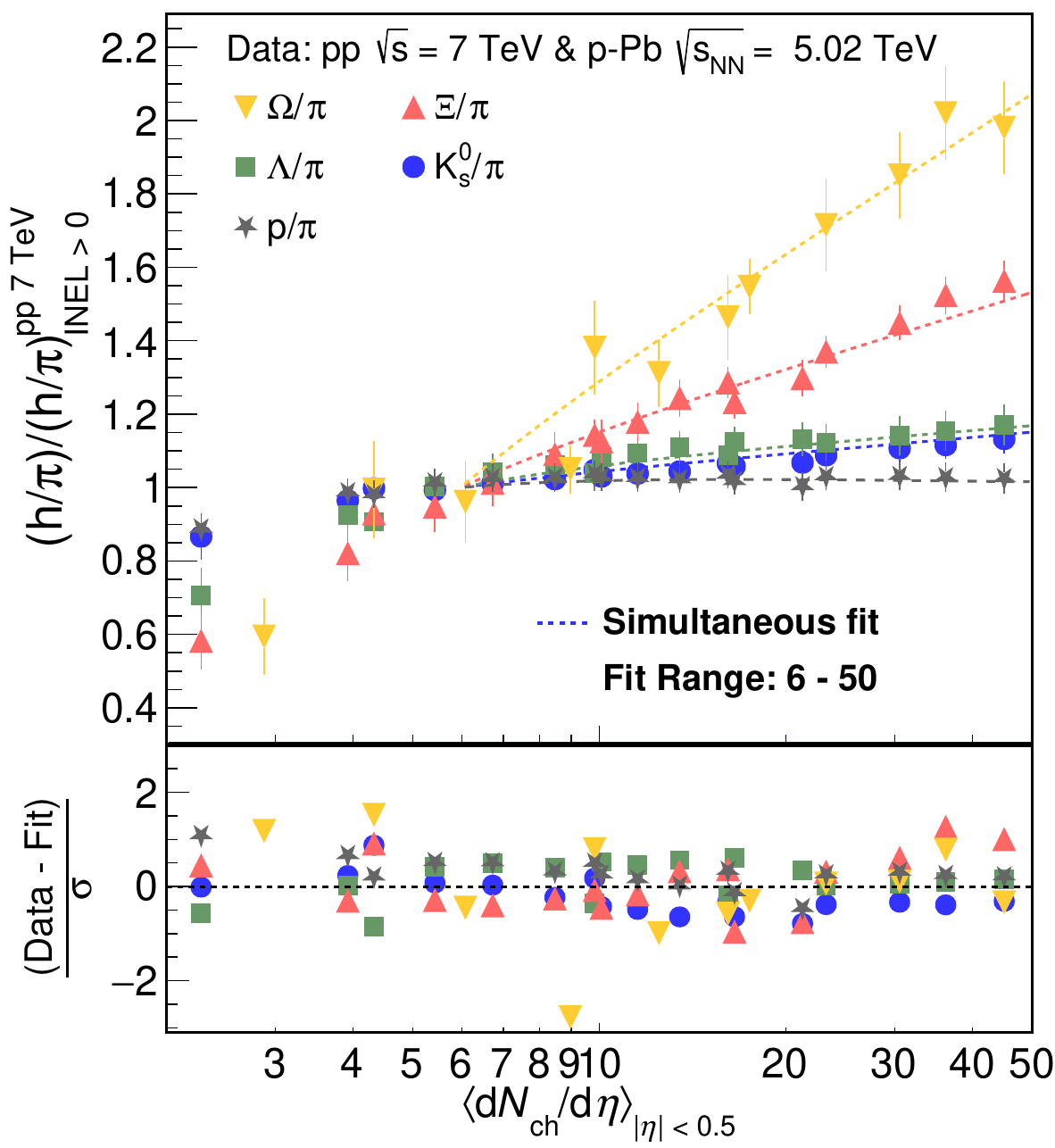}
\caption{Stability test of the present parametrization under variation of the fitted multiplicity interval. The simultaneous fit is performed only within the restricted interval $6 \leq \langle dN_{\mathrm{ch}}/d\eta \rangle \leq 50$, while the resulting parametrization is extrapolated outside the fitted region. The upper panel shows the normalized hadron-to-pion ratios for $K^{0}_{S}/\pi$, $p/\pi$, $\Lambda/\pi$, $\Xi/\pi$, and $\Omega/\pi$, and the lower panel presents the corresponding standardized residuals $(\mathrm{Data}-\mathrm{Fit})/\sigma$.}
\label{fig:FitRange}
\end{figure}
\begin{figure}[thb]
\centering
\includegraphics[width=0.4\columnwidth]{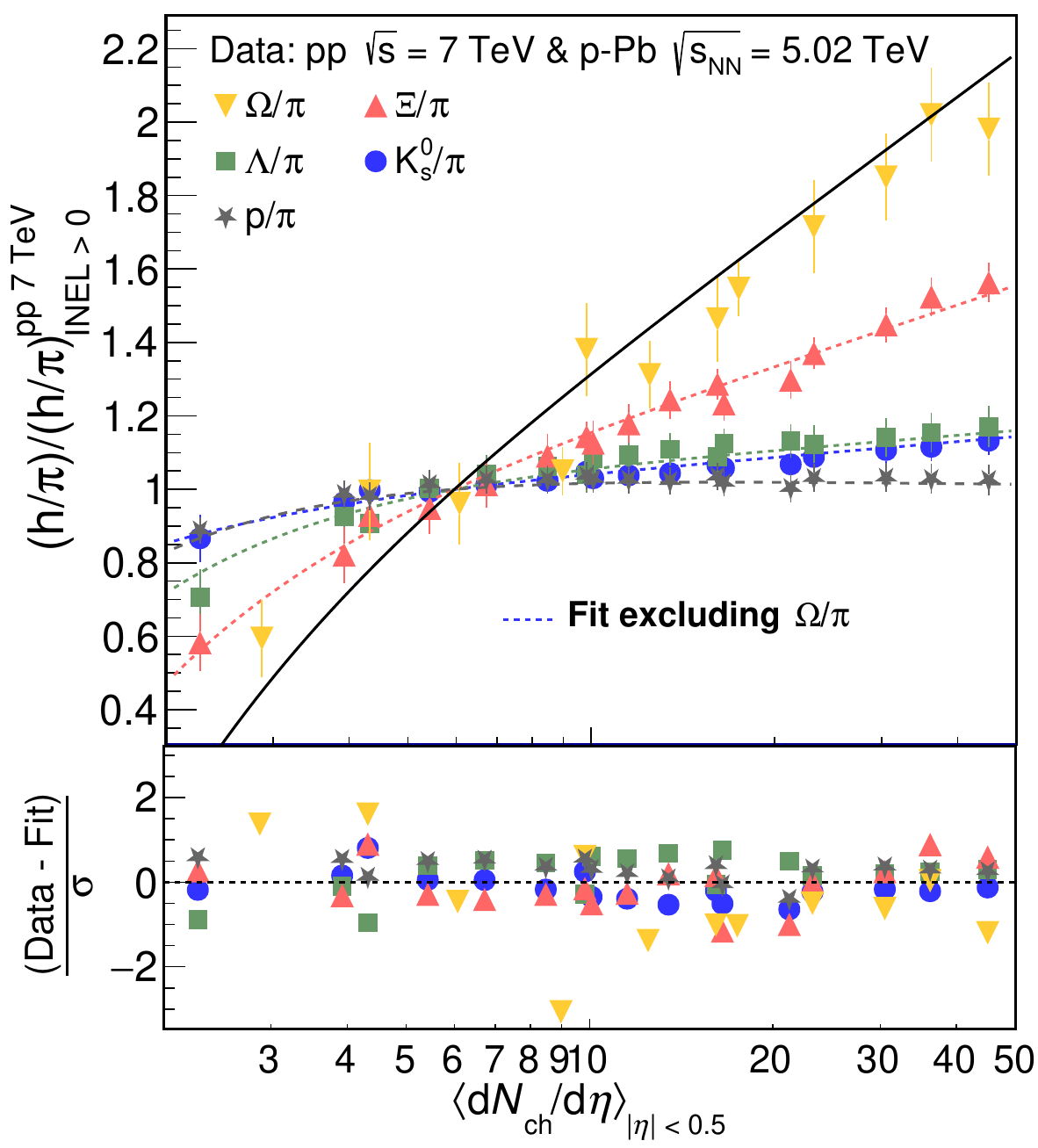}
\caption{Leave-one-out stability test of the present effective mass-scale parametrization. The simultaneous fit is performed excluding the $\Omega/\pi$ observable, while the resulting parametrization is used to describe the measured multiplicity dependence of all hadron-to-pion ratios, including $\Omega/\pi$. The upper panel shows the normalized ratios $(h/\pi)/(h/\pi)_{\mathrm{INEL}>0}^{pp~7~\mathrm{TeV}}$, and the lower panel presents the corresponding standardized residuals $(\mathrm{Data}-\mathrm{Fit})/\sigma$. The reasonable reproduction of the $\Omega/\pi$ trend indicates that the extracted multiplicity hierarchy is not driven solely by the highest-strangeness observable.}
\label{fig:OmegaExcluded}
\end{figure}

Figure~\ref{fig:pPbToPP} presents a cross-system test in which the parameters are determined exclusively from the p--Pb data and subsequently used to predict the multiplicity dependence of the corresponding pp observables. Despite the substantial difference in collision systems and multiplicity coverage, the parametrization reproduces the overall hierarchy and evolution of the measured hadron-to-pion ratios with reasonable consistency. The residual distributions remain approximately centered around zero without strong systematic drift, indicating that the extracted effective mass-scale structure is not restricted to a single collision system.

The stability of the parametrization against variations of the fitted multiplicity interval is examined in Fig.~\ref{fig:FitRange}. In this case, the simultaneous fit is performed only within the restricted interval %
$6 \leq \langle dN_{\mathrm{ch}}/d\eta \rangle \leq 50$, %
while the resulting parametrization is extrapolated outside the fitted region. The observed multiplicity dependence remains comparatively stable over the full measured range, suggesting that the extracted scaling behavior is not driven solely by the low-multiplicity region. In particular, the relative ordering of the strange and multi-strange hadrons is preserved under the modified fit interval.

An additional stability test is shown in Fig.~\ref{fig:OmegaExcluded}, where the $\Omega/\pi$ observable is excluded entirely from the fitting procedure. Since the $\Omega$ baryon corresponds to the highest open-strangeness and largest effective mass scale among the fitted hadrons, this represents a particularly nontrivial leave-one-out test of the framework. The resulting parametrization continues to reproduce the measured $\Omega/\pi$ trend with reasonable agreement, indicating that the extracted hierarchy is not determined solely by the $\Omega$ data points themselves.

Table~IV summarizes the fitted parameters obtained from the baseline analysis and from the various robustness studies discussed above. Although some variation of the parameters is observed under changes of collision system, multiplicity range, and observable selection, the extracted values remain mutually consistent within uncertainties. The corresponding fit qualities also remain comparable across the different tests, indicating that the overall phenomenological description is robust against reasonable variations of the fitting procedure.  These observations suggest that the extracted scaling structure is not driven by a specific subset of observables or by a particular fitting configuration. Taken together, these studies indicate that the present effective mass-scale parametrization provides a comparatively stable phenomenological organization under changes of collision system, fit interval, and observable selection. The observed multiplicity hierarchy therefore appears to reflect broader systematic trends in the data rather than being an artifact of isolated fit constraints or a particular subset of hadronic species.
\begin{table}[t]
\centering
\caption{
Best-fit parameters of the present effective mass-scale parametrization obtained under different robustness and stability tests. The quoted uncertainties correspond to the statistical fit uncertainties returned by the minimization procedure.
}
\label{tab:RobustnessParameters}
\begin{tabular}{lcccc}
\hline\hline
Fit configuration & $a$ & $b$ & $c$ & $\chi^{2}/\mathrm{ndf}$ \\
\hline

Baseline simultaneous fit
& $1.350 \pm 0.134$
& $1.836 \pm 0.107$
& $0.181 \pm 0.035$
& $29.10/76$ \\

Fit to p--Pb data
& $1.483 \pm 0.170$
& $1.906 \pm 0.131$
& $0.207 \pm 0.112$
& $10.03/32$ \\

Fit range: $6 \leq \langle dN_{\mathrm{ch}}/d\eta \rangle \leq 50$
& $1.384 \pm 0.142$
& $1.877 \pm 0.117$
& $0.222 \pm 0.094$
& $21.71/59$ \\

Fit excluding $\Omega/\pi$
& $1.671 \pm 0.314$
& $2.026 \pm 0.179$
& $0.196 \pm 0.036$
& $14.62/65$ \\

\hline\hline
\end{tabular}
\end{table}

\subsection{Disentangling hadron mass and strangeness effects}

A central question in the interpretation of multiplicity-dependent hadron production is whether the observed hierarchy is uniquely determined by open-strangeness content or whether additional effective mass-scale organization is present. Since hadron mass and strangeness are strongly correlated for many identified hadrons, separating these effects experimentally is nontrivial. In the present analysis, this issue is examined through complementary studies of relative enhancement patterns, hidden-strangeness production, and the separate contributions of quark-level and hadronic mass scales within the phenomenological parametrization.

\begin{figure}[htb]
\centering
\includegraphics[width=0.42\columnwidth]{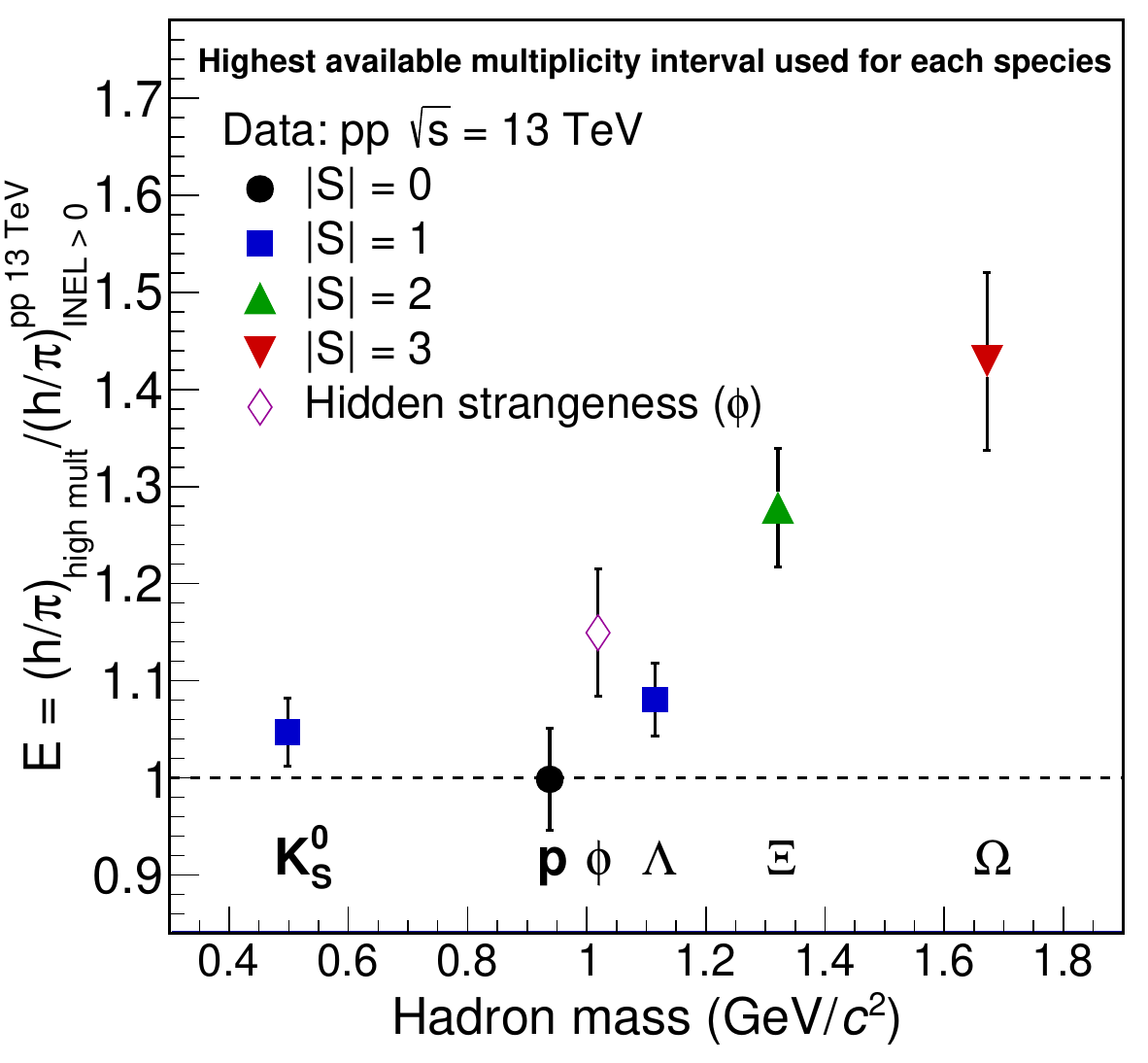}
\caption{
Relative enhancement
$E=(h/\pi)_{\mathrm{high~mult}}/(h/\pi)_{\mathrm{INEL}>0}$
for different hadron species as a function of hadron mass using pp collision data at $\sqrt{s}=13$~TeV. The symbols are grouped according to the magnitude of open strangeness $|S|$, while the hidden-strangeness $\phi$ meson is shown separately. For each species, the highest available multiplicity interval is used. The observed enhancement pattern exhibits a broad mass-related organization with visible species-dependent deviations associated with strangeness structure. The error bars represent the quadratic sum of statistical and uncorrelated systematic uncertainties.
}
\label{fig:EnhancementMass}
\end{figure}

Figure~\ref{fig:EnhancementMass} shows the relative enhancement
\begin{equation}
E=\frac{(h/\pi)_{\mathrm{high~mult}}}{(h/\pi)_{\mathrm{INEL}>0}}
\nonumber
\end{equation}
for different hadron species as a function of hadron mass using pp collision data at $\sqrt{s}=13$~TeV. The pp collision data at $\sqrt{s}=13$~TeV are used for this study because they provide the widest available multiplicity coverage together with measurements of multiple strange and multi-strange hadron species within a common experimental framework. The symbols are grouped according to the magnitude of open strangeness $|S|$, while the $\phi$ meson is shown separately as a hidden-strangeness state. The enhancement exhibits an overall broad mass-related organization with visible species-dependent deviations from $K^{0}_{S}$ to $\Omega$, indicating that the multiplicity hierarchy is not organized solely through integer strangeness counting. In particular, hadrons with similar or identical open-strangeness content do not exhibit identical enhancement behavior. The $K^{0}_{S}$ meson and $\Lambda$ baryon, both corresponding to $|S|=1$, exhibit enhancement values that are broadly compatible within uncertainties despite their different hadronic structure and mass scales, while the hidden-strangeness $\phi$ meson exhibits enhancement comparable to or larger than several open-flavor strange hadrons in a similar mass region.

The comparison between $p$, $\Lambda$, and $\phi$ is especially instructive. These hadrons occupy a relatively similar mass range but possess substantially different valence-quark and strangeness structures. If the multiplicity hierarchy were governed solely by open-strangeness counting, a simpler ordering pattern would be expected. Instead, the observed enhancement indicates a more complex phenomenological organization involving both hadronic and quark-level mass scales.  The present results therefore suggest that the observed hierarchy is not uniquely reducible to simple open-strangeness ordering alone. At the same time, the number of identified hadron species currently available for such comparisons remains limited. Figure~\ref{fig:EnhancementMass} should therefore be interpreted as an illustrative phenomenological trend rather than as a quantitative demonstration of independent mass scaling.

\begin{figure}[t]
\centering
\includegraphics[width=0.48\columnwidth]{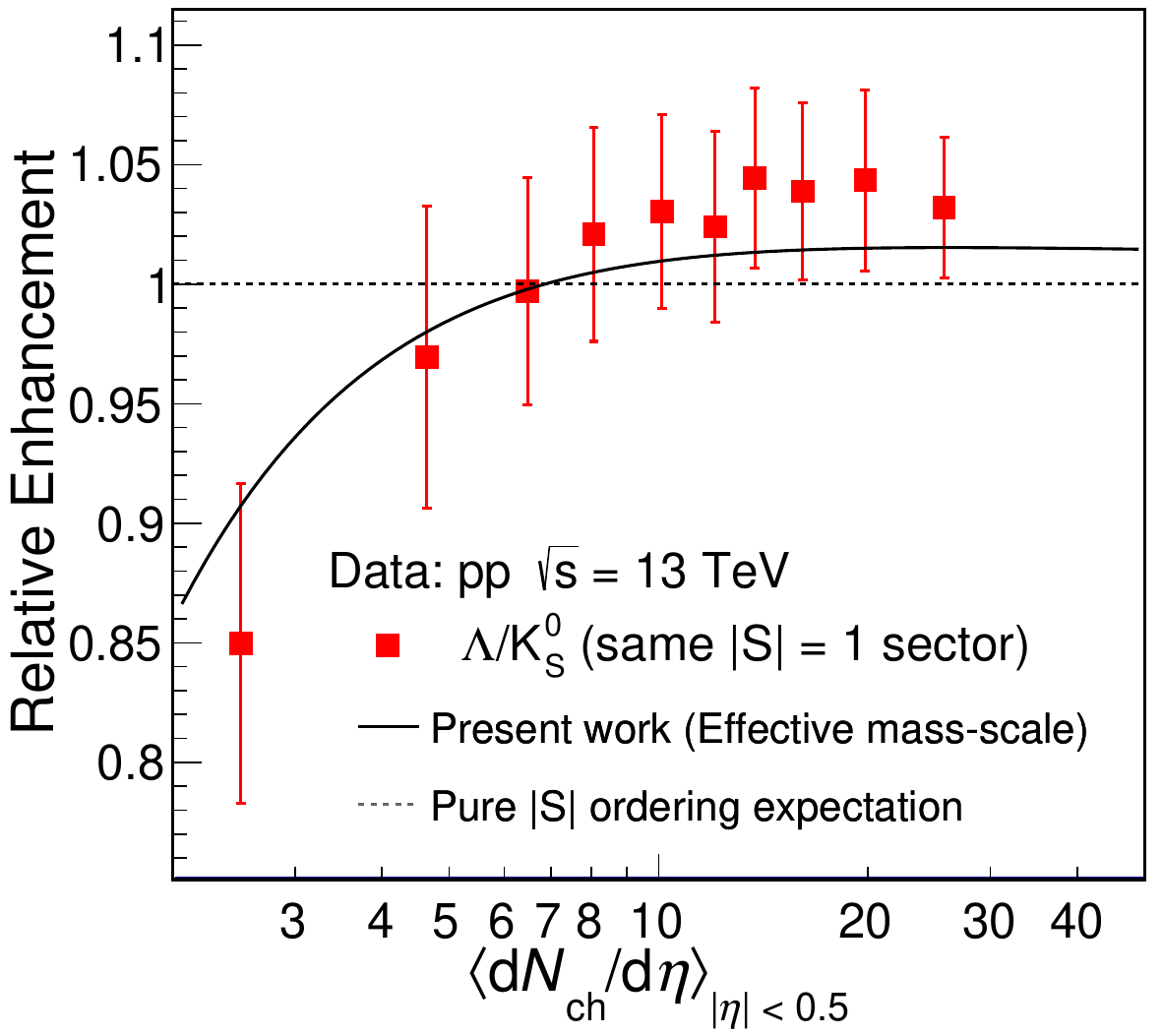}
\caption{
Relative enhancement of the $\Lambda/K_{S}^{0}$ ratio as a function of charged-particle multiplicity in pp collisions at $\sqrt{s}=13$~TeV. Since both hadrons belong to the same open-strangeness sector ($|S|=1$), a strictly identical multiplicity scaling behavior would correspond to a constant ratio near unity. The observed residual multiplicity dependence indicates that hadron-species structure and effective mass-scale effects contribute beyond simple open-strangeness ordering alone. The error bars represent the quadratic sum of statistical and uncorrelated systematic uncertainties.
}
\label{fig:LamKs0}
\end{figure}

To further examine the extent to which the observed multiplicity hierarchy can be reduced to simple open-strangeness ordering, Fig.~\ref{fig:LamKs0} shows the relative enhancement of the $\Lambda/K_{S}^{0}$ ratio as a function of charged-particle multiplicity in pp collisions at $\sqrt{s}=13$~TeV. Since both $\Lambda$ and $K_{S}^{0}$ belong to the same open-strangeness sector ($|S|=1$), a strictly identical multiplicity scaling behavior would lead to an approximately constant ratio close to unity over the full multiplicity range. Instead, the measured ratio exhibits a mild but systematic multiplicity dependence. The present parametrization reproduces this behavior consistently, indicating that hadrons within the same open-strangeness sector do not necessarily follow completely identical scaling trends. This suggests that additional hadron-species-dependent effects appear to contribute beyond a purely strangeness-based ordering. In particular, baryon--meson differences and effective mass-scale contributions appear to influence the residual multiplicity evolution even within the same open-strangeness sector. At the same time, the observed deviation from unity remains moderate, indicating that open strangeness continues to play an important organizing role in the overall enhancement hierarchy. The present results therefore support the interpretation that the multiplicity dependence of identified hadron production reflects an interplay between strangeness content, hadron-species structure, and effective mass-scale organization rather than being uniquely determined by a single variable alone.

\begin{figure}[htb]
\centering
\includegraphics[width=0.42\columnwidth]{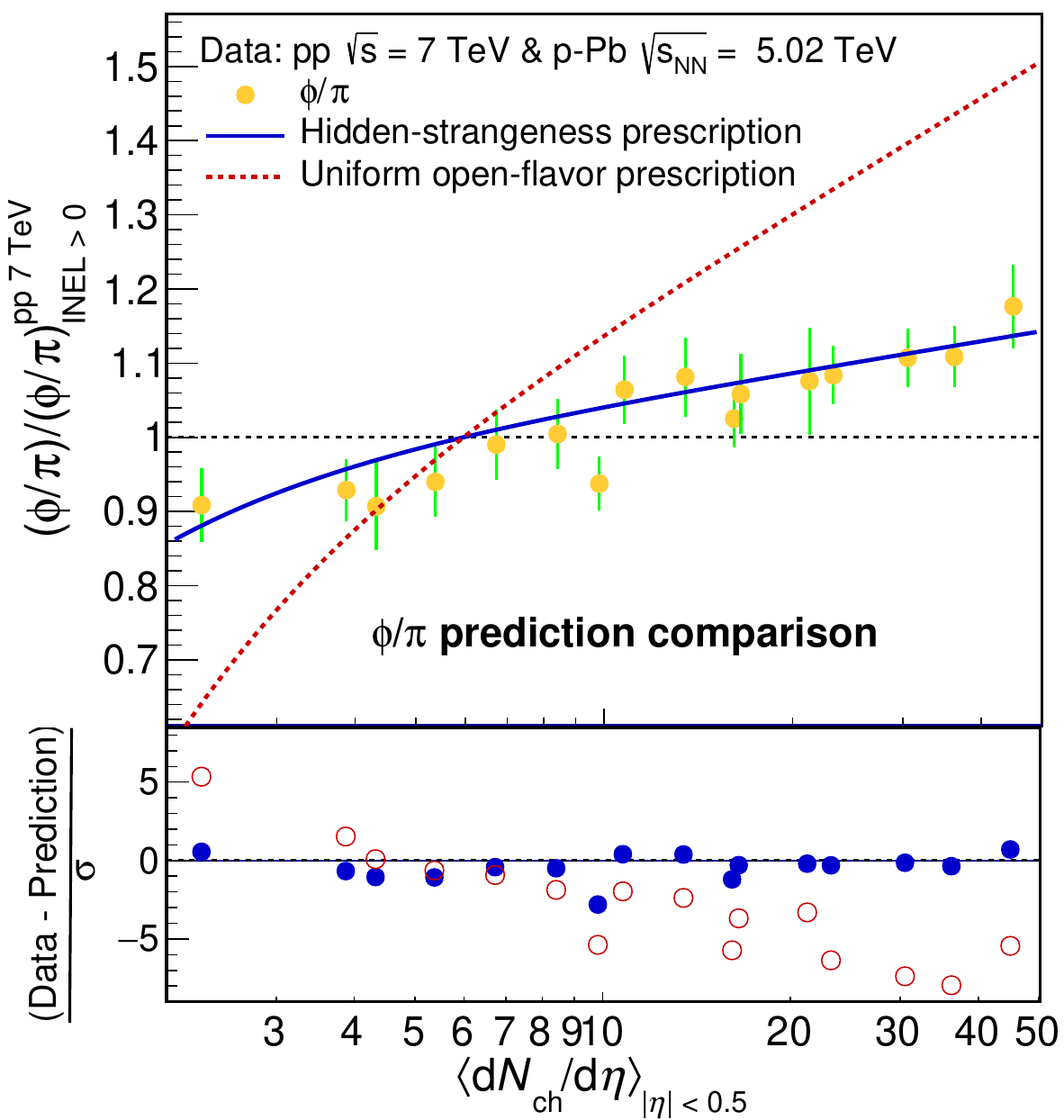}
\caption{
Comparison of the normalized $\phi/\pi$ multiplicity dependence with two different effective mass-scale prescriptions for the $\phi$ meson. The upper panel shows the measured ratio $(\phi/\pi)/(\phi/\pi)_{\mathrm{INEL}>0}^{pp~7~\mathrm{TeV}}$ together with the predictions obtained using the hidden-strangeness and the uniform open-flavor prescriptions defined in Sec.~II. The lower panel presents the corresponding standardized residuals $(\mathrm{Data}-\mathrm{Prediction})/\sigma$. All curves are obtained using the same parameter set determined from the simultaneous fit discussed previously, without any additional refitting for the $\phi/\pi$ observable. The hidden-strangeness prescription provides a significantly more consistent description of the observed multiplicity dependence.
}
\label{fig:PhiComparison}
\end{figure}

An additional probe of the interplay between mass-scale organization and strangeness structure is provided by the hidden-strangeness $\phi$ meson. In the present framework, the $\phi$ meson is treated using a hidden-strangeness prescription that differs from the uniform open-flavor scaling applied to the other hadrons. This distinction is motivated by the hidden-strangeness $(s\bar{s})$ structure of the $\phi$ meson and is examined here as a prediction-level test rather than as an additional fit constraint. The hidden-strangeness prescription is motivated by the absence of net open strange transport in the $\phi$ meson and its hidden-strangeness $(s\bar{s})$ structure, rather than being introduced as an empirical tuning prescription. Figure~\ref{fig:PhiComparison} compares the measured multiplicity dependence of the normalized $\phi/\pi$ ratio with the predictions obtained using the two prescriptions. The hidden-strangeness prescription provides a noticeably improved description of the observed multiplicity evolution, while the uniform open-flavor treatment systematically overestimates the enhancement at intermediate and high multiplicities. This behavior is also reflected in the standardized residual distributions shown in the lower panel. The $\phi$ meson therefore constitutes a nontrivial probe of the proposed phenomenological framework and indicates that hidden-strangeness states cannot be incorporated through a naive extension of the open-flavor scaling prescription. The present treatment should not be interpreted as a microscopic model of $\phi$ production, but rather as a phenomenological test of whether hidden-strangeness states can be incorporated consistently within the proposed effective scaling framework.

\begin{figure}[htb]
\centering
\includegraphics[width=0.40\columnwidth]{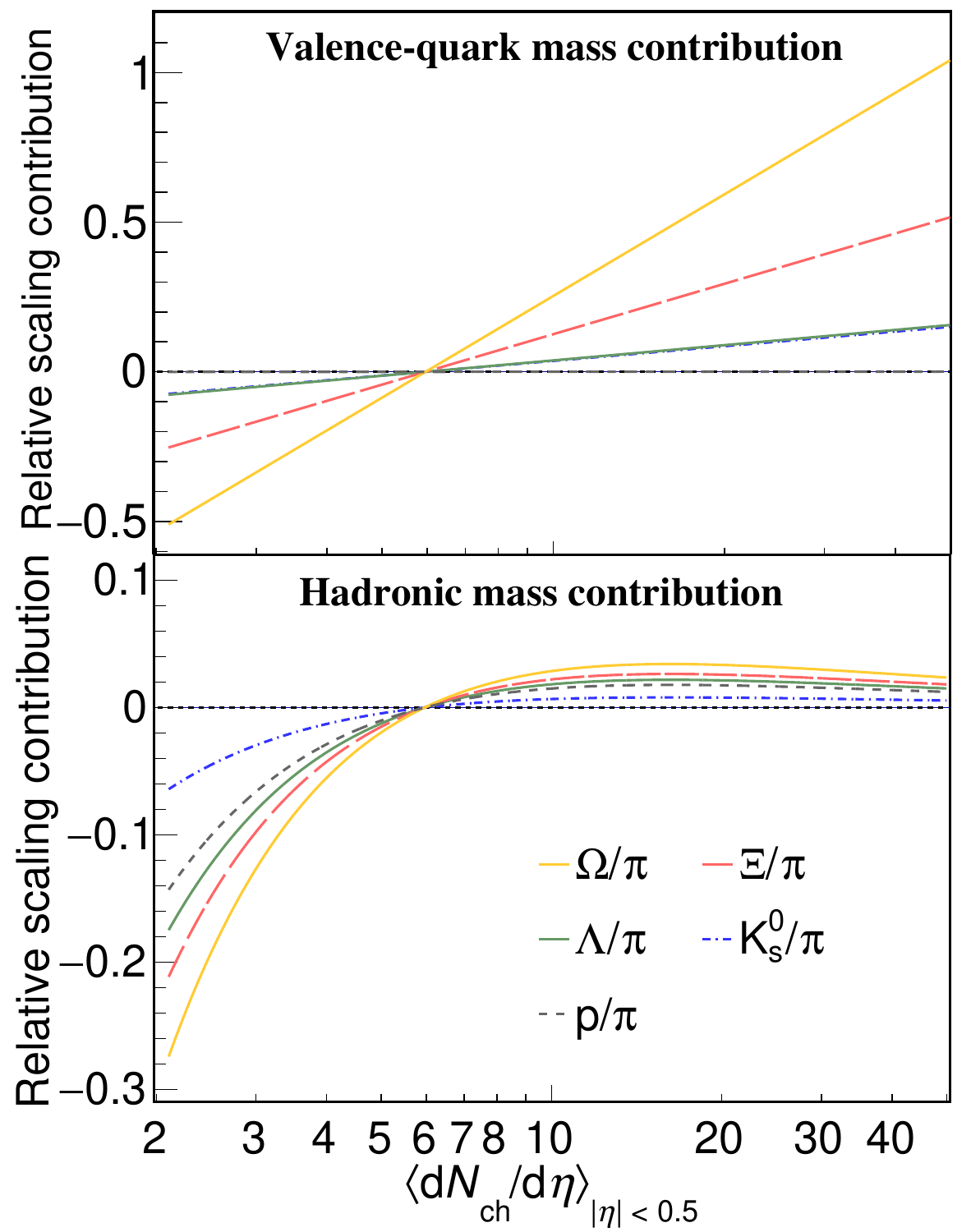}
\caption{
Separate contributions associated with the effective valence-quark mass term (upper panel) and hadronic mass term (lower panel) entering the phenomenological parametrization of Eq.~(\ref{eq:parametrization}). The decomposition illustrates that the dominant multiplicity hierarchy is primarily associated with the effective quark-level mass contribution, while the hadronic mass term provides a smaller but non-negligible correction, particularly for heavier strange hadrons and lower multiplicities.
}
\label{fig:ContributionDecomposition}
\end{figure}

To further examine the phenomenological structure of the parametrization, Fig.~\ref{fig:ContributionDecomposition} shows the separate contributions associated with the valence-quark mass term and the hadronic mass term entering Eq.~(\ref{eq:parametrization}). The decomposition indicates that the dominant ordering of the multiplicity hierarchy originates primarily from the effective valence-quark mass contribution, while the hadronic mass term provides a smaller but non-negligible correction. The hadronic contribution becomes relatively more visible at lower multiplicities and for heavier strange hadrons, indicating that both components contribute to the overall organization of the observed enhancement hierarchy.
 
It should be emphasized that these contributions should not be interpreted as independently measurable physical components of hadron production. Rather, they represent a decomposition within the present phenomenological parametrization intended to illustrate the relative roles of effective quark-level and hadronic mass scales in organizing the observed multiplicity dependence.

Taken together, these observations indicate that the observed enhancement hierarchy appears to reflect an interplay between open strangeness, hadron species, and effective mass-related scales, rather than being uniquely characterized by simple integer strangeness ordering alone. The present analysis does not imply that strangeness is irrelevant in hadron production, nor does it suggest that hadron mass alone constitutes a unique organizing variable. Rather, the results support the interpretation that the multiplicity dependence of identified hadron production admits a broader phenomenological organization involving correlated contributions from hadronic structure, hidden-strangeness effects, and effective mass-scale behavior within the proposed framework.

\section{Physical interpretation and limitations}

The present framework should be interpreted as an effective phenomenological organization of the observed multiplicity hierarchy rather than as a microscopic derivation of hadron production dynamics. The parametrization is motivated by the observation that strange and multi-strange hadron production is intrinsically correlated with increasing hadronic and valence-quark mass scales. Since heavier identified hadrons in the light-flavor sector typically contain one or more strange quarks, multiplicity-dependent reduction of suppression effects associated with strange-quark production can naturally generate correlated scaling behavior involving both strangeness and effective mass-related quantities.

One possible interpretation is that the multiplicity dependence reflects effective threshold-like production scales associated with heavier hadrons and strange-quark pair creation. In low-multiplicity environments, strange and multi-strange hadrons are comparatively suppressed, while increasing event activity gradually reduces these suppression effects, leading to the observed enhancement hierarchy. Within such a picture, hadron production may organize phenomenologically according to effective quark-level and hadronic mass scales that are correlated with the underlying production thresholds.

Related correlations between hadron production and effective quark-level scales also appear in several existing approaches to small-system hadron production, although the underlying microscopic mechanisms differ substantially. For example, canonical suppression models associate strange-hadron enhancement with the gradual relaxation of local conservation constraints in larger effective production volumes. Likewise, recombination/coalescence approaches and string-based hadronization scenarios, including rope-hadronization models, naturally generate enhanced strange and multi-strange hadron production in environments with increased partonic density or string overlap. The present framework does not attempt to reproduce these microscopic mechanisms explicitly, but instead provides a compact phenomenological organization of the resulting multiplicity hierarchy.

The use of current quark masses in the present framework should not be interpreted literally as implying perturbative quark masses governing hadronization dynamics. Rather, the quark-mass quantities employed here act as effective ordering variables within the phenomenological scaling organization explored in the present analysis. In this sense, the parametrization is intended to probe correlated scaling behavior involving hadronic and quark-level mass-related quantities rather than to establish a microscopic constituent description of hadron formation. We have verified that the qualitative hierarchy obtained in the present analysis is not sensitive to the precise numerical choice of the light- and strange-quark masses. Since the parametrization depends primarily on the relative ordering of the effective quark-mass scales rather than on their absolute numerical values, the use of current quark masses should be regarded as a convenient phenomenological convention rather than a unique physical choice.

The disentangling studies performed in the present work further indicate that the observed enhancement hierarchy is not uniquely reducible to simple open-strangeness counting alone. In particular, the hidden-strangeness $\phi$ meson provides a nontrivial test of the proposed scaling structure, while the enhancement patterns observed in the similar-mass region involving $p$, $\Lambda$, and $\phi$ suggest that hadron species dependence and effective mass-related quantities contribute simultaneously to the observed organization. At the same time, the present analysis does not imply that hadron mass itself constitutes the unique or fundamental dynamical variable governing hadron production. Rather, the results suggest that effective mass-scale organization provides a complementary phenomenological perspective on the multiplicity dependence of identified hadron production.

Several limitations of the present framework should also be emphasized. The parametrization is not expected to provide a universal description of all hadronic observables or collision systems without modification. In particular, heavy-flavor hadron production involves additional hard-scattering and heavy-quark dynamical mechanisms that are beyond the scope of the present analysis. Likewise, dense Pb--Pb collision systems may contain substantial collective, medium-induced, and rescattering effects that are not incorporated in the present effective scaling framework. Resonance observables with strong hadronic re-scattering contributions may also require additional dynamical ingredients beyond the present parametrization. Consequently, the present study should be viewed as a phenomenological scaling analysis applicable primarily to identified light-flavor hadron production in small collision systems at the LHC. It should also be emphasized that the present parametrization is not necessarily unique, and alternative effective scaling forms may provide comparably successful phenomenological descriptions of the data. The primary purpose of the present framework is therefore not to establish a unique microscopic interpretation of hadron production, but rather to examine whether the observed multiplicity hierarchy admits a complementary phenomenological organization involving effective hadronic and valence-quark mass scales.

\section{Conclusions}

In this work, we have studied the multiplicity dependence of identified hadron yield ratios in high-energy $pp$ and $p$--Pb collisions within a phenomenological effective mass-scale framework. The central aim of the analysis was to examine whether the observed multiplicity hierarchy, which is commonly discussed in terms of strangeness-driven scaling and canonical suppression, may also admit a complementary organization through effective hadronic and valence-quark mass scales relative to the pion baseline. A simultaneous fit to a broad set of hadron-to-pion ratios, including non-strange, strange, and multi-strange hadrons, shows that the present parametrization provides a consistent description of the measured data over the available multiplicity range.

An important outcome of the analysis is the predictive capability of the parametrization. Using parameters fixed only from the simultaneous fit to the $h/\pi$ ratios, the framework describes several independent observables that were not included in the fit, including ratios with different denominators and resonance production. In particular, the successful description of observables such as $\Xi/K^{0}_{S}$, $\Lambda/K^{0}_{S}$, $p/K^{0}_{S}$, $\phi/\pi$, and $\Xi^{*}/\pi$ indicates that the parametrization captures systematics beyond the specific data used to determine the fit parameters. The robustness tests further show that the extracted scaling structure remains comparatively stable under changes in the collision system used for the fit, variations of the fitted multiplicity interval, and the exclusion of selected hadron species such as $\Omega/\pi$. This stability suggests that the observed hierarchy is not driven by an isolated subset of data points or by a particular fit choice.

The disentangling studies provide further insight into the underlying physics. The enhancement-vs-mass analysis in pp collisions at $\sqrt{s}=13$~TeV shows that the multiplicity hierarchy is not uniquely determined by simple open-strangeness counting alone. Hadron species with similar open-strangeness content exhibit visible but moderate species-dependent deviations from a purely universal scaling trend, and the hidden-strangeness $\phi$ meson emerges as a particularly sensitive probe of the proposed scaling framework. The dedicated comparison of two prescriptions for the $\phi/\pi$ ratio shows that the hidden-strangeness treatment gives a markedly more consistent description than a uniform open-flavor prescription. The decomposition of the parametrization indicates that the effective valence-quark mass term provides the leading phenomenological component of the observed scaling behavior, while the hadronic mass contribution remains smaller but non-negligible, especially for heavier hadrons and at lower multiplicities.

At the same time, the present analysis should be viewed as a phenomenological scaling study rather than a microscopic theory of hadron production. The results do not imply that strangeness is unimportant, nor do they replace thermal, canonical suppression, or microscopic dynamical interpretations. Rather, they indicate that the multiplicity dependence of hadron production may admit a broader effective organization involving an interplay of strangeness, hadron species dependence, hidden-strangeness structure, and effective mass-related scales. The present parametrization is not necessarily unique, and alternative phenomenological scaling forms may also provide comparably successful descriptions of the data. The framework is therefore intended as a complementary description that probes whether the observed hierarchy can be organized within an effective mass-scale perspective, while remaining consistent with existing strangeness-based and thermal interpretations. Future measurements involving charm hadrons, resonances with strong hadronic rescattering contributions, and heavy-ion collision systems will provide additional tests of the extent to which the proposed effective mass-scale organization remains applicable.


\bibliographystyle{apsrev4-2} 
\bibliography{references}      

\end{document}